\documentclass[reprint,twocolumn,aps,prB,amsmath,amssymb,floatfix,superscriptaddress,longbibliography]{revtex4-1}
\usepackage{graphicx}
\usepackage{epsfig}
\usepackage{bm}
\usepackage{dcolumn}
\usepackage{color}
\usepackage{physics}
\usepackage{float}
\usepackage[colorlinks,urlcolor=cyan,citecolor=blue,linkcolor=magenta]{hyperref}
\makeatletter
\newcommand*{\rom}[1]{\expandafter\@slowromancap\romannumeral #1@}
\makeatother
\usepackage{tikz}
\usepackage{subfigure}

\begin{document}

\preprint{APS/123-QED}
\preprint{This line only printed with preprint option}

\title{Asymmetric transfer matrix analysis of Lyapunov exponents in one-dimensional non-reciprocal quasicrystals}

\author{Shan-Zhong Li}
\affiliation {Key Laboratory of Atomic and Subatomic Structure and Quantum Control (Ministry of Education), Guangdong Basic Research Center of Excellence for Structure and Fundamental Interactions of Matter, School of Physics, South China Normal University, Guangzhou 510006, China}

\affiliation {Guangdong Provincial Key Laboratory of Quantum Engineering and Quantum Materials, Guangdong-Hong Kong Joint Laboratory of Quantum Matter, Frontier Research Institute for Physics, South China Normal University, Guangzhou 510006, China}

\author{Enhong Cheng}
\affiliation {Key Laboratory of Atomic and Subatomic Structure and Quantum Control (Ministry of Education), Guangdong Basic Research Center of Excellence for Structure and Fundamental Interactions of Matter, School of Physics, South China Normal University, Guangzhou 510006, China}

\affiliation {Guangdong Provincial Key Laboratory of Quantum Engineering and Quantum Materials, Guangdong-Hong Kong Joint Laboratory of Quantum Matter, Frontier Research Institute for Physics, South China Normal University, Guangzhou 510006, China}

\author{Shi-Liang Zhu}
\affiliation {Key Laboratory of Atomic and Subatomic Structure and Quantum Control (Ministry of Education), Guangdong Basic Research Center of Excellence for Structure and Fundamental Interactions of Matter, School of Physics, South China Normal University, Guangzhou 510006, China}

\affiliation {Guangdong Provincial Key Laboratory of Quantum Engineering and Quantum Materials, Guangdong-Hong Kong Joint Laboratory of Quantum Matter, Frontier Research Institute for Physics, South China Normal University, Guangzhou 510006, China}
\affiliation{Quantum Science Center of Guangdong-Hong Kong-Macao Greater Bay Area, Shenzhen, China}

\author{Zhi Li}
\email[Corresponding author: ]{lizphys@m.scnu.edu.cn}
\affiliation {Key Laboratory of Atomic and Subatomic Structure and Quantum Control (Ministry of Education), Guangdong Basic Research Center of Excellence for Structure and Fundamental Interactions of Matter, School of Physics, South China Normal University, Guangzhou 510006, China}

\affiliation {Guangdong Provincial Key Laboratory of Quantum Engineering and Quantum Materials, Guangdong-Hong Kong Joint Laboratory of Quantum Matter, Frontier Research Institute for Physics, South China Normal University, Guangzhou 510006, China}

\date{\today}

\begin{abstract}
The Lyapunov exponent, serving as an indicator of the localized state, is commonly utilized to identify localization transitions in disordered systems. In non-Hermitian quasicrystals, the non-Hermitian effect induced by non-reciprocal hopping can lead to the manifestation of two distinct Lyapunov exponents on opposite sides of the localization center. Building on this observation, we here introduce a comprehensive approach for examining the localization characteristics and mobility edges of non-reciprocal quasicrystals, referred to as asymmetric transfer matrix analysis. We demonstrate the application of this method to three specific scenarios: the non-reciprocal Aubry-Andr\'e model, the non-reciprocal off-diagonal Aubry-Andr\'e model, and the non-reciprocal mosaic quasicrystals. This work may contribute valuable insights to the investigation of non-Hermitian quasicrystal and disordered systems.
\end{abstract}

\maketitle

\section{Introduction}
Anderson localization is a fundamental concept in condensed matter physics, elucidating particles exponentially localized due to random disorder~\cite{PWAnderson1958}. In one-dimensional (1D) and two-dimensional (2D) systems, arbitrarily small disorder leads to all eigenstates being localized~\cite{EAbrahams1979,PALee1985,BHetenyi2021}. In  three-dimensional  systems, below a specific disorder strength threshold, extended and localized states can coexist, separated by mobility edges~\cite{FEvers2008,ALagendijk2009}. In contrast to random disorder, quasiperiodic systems offer an experimental platform for investigating Anderson localization and mobility edges in low dimensions~\cite{LFallani2007,GRoati2008,YLahini2009,DTanese2014,FAAn2018,VGoblot2020,FAAn2021,YWang2022,HLi2023,HPLuschen2018}. The well-known Aubry-Andr\'e (AA) model, due to its self-duality property, exhibits no mobility edge, with the system being either in an extended or localized state depending on whether the quasiperiodic strength is below or above the critical value, respectively ~\cite{SAubry1980}. Mobility edges can be observed in various generalized versions of the AA model by introducing short-range or long-range hopping interactions~\cite{JBiddle2011,XXia2022,XDeng2019,MGoncalves2023,JBiddle2010,SRoy2021} or by modifying the quasiperiodic potentials~\cite{APadhan2022,SDasSarma1988,SDasSarma1990,TLiu2022,SGaneshan2015,XLi2017,HYao2019,XLi2020,YCZhang2022,XCZhou2023,YWang2020,XPLi2016,ZLu2022}.

The exploration of non-Hermitian systems has garnered significant attention in recent years, leading to the discovery of many phenomena that are absent in Hermitian systems~\cite{ZWang2022,DWZhang2020,VVKonotop2016,RElGanniny2018,YAshida2020,CMBender1998,AGuo2009,ARegensburger2012,SWeimann2017,MKremer2019,SXia2021,YLi2022,PPeng2016,JLi2019,ZRen2022,LXiao2017,LLi2020, EJBergholtz2021,LZhou2023,KLi2023,LZhou2024,SZLi2024,KKawabata2023,HZLi2023,XJYu2023,ZXGuo20222,CHXu2023,JLi2023}. A notable discovery is the non-Hermitian skin effect~\cite{VMMartinez2018,YXiong2018,SYao2018,ZGong2018}, which holds significance in understanding non-Hermitian topological band theory. This effect typically manifests in non-reciprocal hopping systems, where the eigenstates and eigenvalues are highly sensitive to boundary conditions. Specifically, under open boundary conditions (OBC), the eigenstates exhibit exponential localization at the boundaries; On the other hand, under periodic boundary conditions (PBC), the eigenenergies form closed loops in the complex plane, characterized by non-zero spectral winding numbers and topological point gaps that describe the skin effect~\cite{KZhang2020}. This sensitivity to boundary conditions changes the traditional bulk-boundary correspondence~\cite{VMMartinez2018,YXiong2018,ZGong2018,SYao2018,KKawabata2019,CHLee2019,KYokomizo2019,LXiao2020, KZhang2020, NOkuma2020, DSBorgnia2020, ZYang2020, CXGuo2021,LJLang2021}.

Recently, the interplay between non-Hermitian and quasiperiodic systems has suggested the existence of a localization phase or mobility edge characterized by the spectral winding number~\cite{DWZhang2020a,HJiang2019,SLonghi2019,YLiu2020,QBZeng2020a,QBZeng2020b,TLiu2020,XCai2021,YLiu2021a,JClaes2021,LZTang2021,SLonghi2021a,SLonghi2021b,LZhou2022,XCai2022,WHan2022,QLin2022,YLiu2021,LJZhai2020,LZTang2022,TQian2024}. In Hatano-Nelson type quasicrystals under OBCs  (PBCs), a transition from the skin phase (extended phase) to the Anderson localized phase occurs as the quasiperiodic strength increases~\cite{NHatano1996}. This transition is accompanied by a change in the spectral winding number from non-zero to zero~\cite{HJiang2019}. The model can be mapped to reciprocal quasicrystals through a similarity transformation under OBCs, revealing that the system is in the skin phase before a critical point, with the eigenstates exponentially localized at the boundary. In Anderson localization region, the system has  asymmetric localized states with two distinct Lyapunov exponents (LEs)~\cite{HJiang2019,APAcharya2024,LZTang2021,XCai2021,XCai2022,ZHWang2021,SLJiang2023}. Notably, not all systems exhibit the skin effect and the LEs pair of localized states is linked through a similarity transformation. For example, in the disordered Hatano-Nelson model, involving non-reciprocal nearest-neighbor hopping with additional disorder couplings, a modified generalized Brillouin zone is required to manifest the skin effect~\cite{ZQZhang2023,HLiu2023}. However, it remains unclear if this method based on the modified generalized Brillouin zone can effectively describe localized phase transitions and the LEs pair of localized states.

In disordered systems, the LE can be calculated using the transfer matrix method. Avila has recently introduced a comprehensive theoretical framework for exact determination of the LE in single-frequency quasiperiodic systems~\cite{Avila2015}, which has also been extended to non-Hermitian systems more recently~\cite{YLiu2021,YLiu2021a}. In non-reciprocal systems, the presence of two LEs necessitates characterization using two transfer matrices. In this article, we apply transfer matrices on the left and right sides to determine the two LEs of the localized state, enabling an accurate characterization of the localized transition. Initially, we apply this method to systems capable of similar transformations, specifically non-reciprocal AA models. By integrating Avila's global theory, we precisely compute the two LEs of the localized state, generating consistent results with those of similar transformations. Subsequently, we extend this analysis to systems unable to undergo similar transformations, including the non-reciprocal off-diagonal AA model (lacking a mobility edge) and the non-reciprocal mosaic quasicrystal (featuring a mobility edge). By contrasting the two distinct LEs, we discover that apart from localized phase transitions, they can also effectively describe extended-critical phase transitions and topological phase transitions.


The paper is organized as follows. We provide a warm-up on the non-reciprocal AA model in Sec.~\ref{Sec.2}. Then, we apply the transfer matrix method to systems that cannot be described by the similarity transformation method. In Sec.~\ref{Sec.4}, the case of quasicrystals with the non-reciprocal nearest-neighbor quasiperiodic coupling is discussed. In Sec.~\ref{Sec.5}, as a typical example of models with mobility edge, we study non-reciprocal mosaic AA models. Main findings of this paper are concluded in Sec.~\ref{Sec.6}.

\section{Non-Hermitian Aubry-Andr\'e model}\label{Sec.2}

Let's warm up with a non-reciprocal quasicrystal; the corresponding Hamiltonian is given by
\begin{equation}\label{Hami}
H=\sum_{j=1}^{L-1}(t_{l,j}c_{j}^{\dagger}c_{j+1}+t_{r,j}c_{j+1}^{\dagger}c_{j})+\sum_{j=1}^{L}V_{j}c_{j}^{\dagger}c_{j},
\end{equation}
where $c_{j}^{\dagger}$ ($c_{j}$) denotes creation (annihilation) operators on the $j$-th site, $L$ is the total number of lattice sites, $t_{l,j}$ ($t_{r,j}$) represent the strength of the left- (right-) hopping term between sites $j$ and $j+1$. For a general quasiperiodic potential $V_{j}=2\lambda\cos(2\pi\alpha j+\theta)$, it involves the following parameters: $\lambda$ denotes the strength of the quasiperiodic potential, $\alpha$ is the quasiperiodic parameter, and $\theta$ represents the global phase. When $t_{l,j}=t_{r,j}=t$, the system reduces to the case with symmetric hopping, and the localization transition occurs at $\lambda=t$~\cite{SAubry1980}. In numerical calculations, without loss of generality, we set $\theta=0$. The irrational number $\alpha$ can be approximated as $\frac{F_{m}}{F_{m+1}}$ for PBCs and $\alpha=\lim_{m\rightarrow\infty}\frac{F_{m}}{F_{m+1}}=\frac{\sqrt{5}-1}{2}$ for OBCs, where $F_{m}$ denotes the $m$-th Fibonacci number.

We first review the non-reciprocal AA model and its Hamiltonian is given by~\cite{HJiang2019}:
\begin{equation}
t_{l,j}=e^{g},~t_{r,j}=e^{-g},~V_{j}=2\lambda\cos(2\pi\alpha j+\theta).
\end{equation}
To study the localization phase transition of the Hamiltonian Eq.~\eqref{Hami}, we analyze the system by the traditional similarity transformation method and the new asymmetric transfer matrix method. First, let's start at the similarity transformation method.

\subsection{Similarity transformations}
The asymmetric hopping ($g\neq 0$) induces the skin effect for OBCs, and the Hamiltonian can be transformed into the Hermitian AA model using a similarity transformation, and is given by
\begin{equation}
H'=SHS^{-1}=\begin{pmatrix} V_{1} & t &   & \\t &  V_{2} & t & \\   & \ddots  & \ddots  &t \\  &  & t &  V_{L}
\end{pmatrix},
\end{equation}
where $S=\mathrm{diag}(e^{-g},~e^{-2g},~\dots,~e^{-Lg})$ is the similarity matrix. For the Hermitian Hamiltonian $H'$, the localization transition point is $\lambda=t$ with $t=\sqrt{t_{l}t_{r}}=1$. Let $\psi'$ be an eigenstate of the Hamiltonian $H'$; then, the eigenstate $\psi$ of the Hamiltonian $H$ satisfies $\psi=S^{-1}\psi'$. Thus, for an extended eigenstate of the Hamiltonian $H'$, $S^{-1}$ causes the wave function to be exponentially localized on the left (right) boundary for $g>0$ ($g<0$), which results in non-Hermitian skin effects. For localized state, the wave function will decay exponentially, that is, the relationship
\begin{equation}\label{fit}    
|\psi_{j}|\propto \left\{\begin{matrix}e^{-\gamma_{r}(j-j_{0})},& j>j_{0}, \\e^{-\gamma_{l}(j_{0}-j)},& j<j_{0},\end{matrix}\right.
\end{equation}
is satisfied, where $j_{0}$ is the site index of the localization center. The expression for the LEs to the left and right of site $j_0$:
\begin{equation}\label{gammanrAA}\begin{aligned}&\gamma_{l}=\eta-g ~~\&~~ \gamma_{r}=\eta+g\end{aligned}
\end{equation}
where $\eta = \max\left\{\ln(\frac{\lambda}{t}),0\right\}$ is the LE for Hamiltonian $H'$. Thus, the eigenstate $\psi$ is asymmetric with two LEs for $g>0$. In the case where $\eta\le |g|$, the system is in the skin phase, and the eigenstate is localized at the left boundary. The skin-Anderson phase transition point is determined by $\lambda=\max\left\{t_{l},t_{r}\right\}$ when $\gamma_{l},\gamma_r>0$. The localization transition points for PBCs and the asymmetric localized states are the same as for OBCs, which all rely on the system being similarly transformed~\cite{APAcharya2024,XCai2021,YLiu2021}.

\begin{figure}[htbp]
\centering
\includegraphics[width=8.5cm]{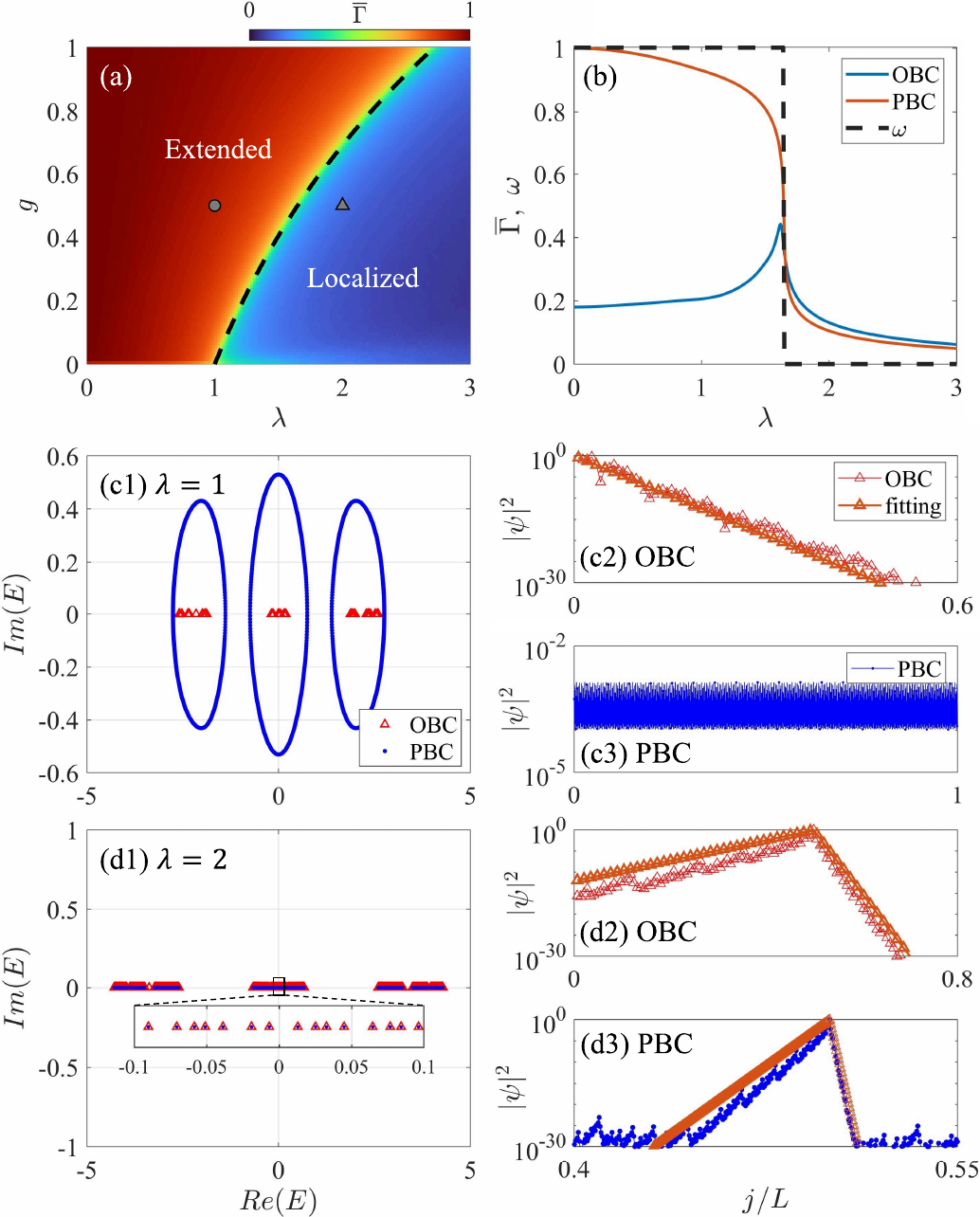}
\caption{(a) The average fractal dimension $\overline{\Gamma}$ in $\lambda-g$ plane for the system size $L=610$, where circles and triangles represent the positions of $\lambda=1$ and $\lambda=2$ at $g=0.5$, respectively. (b) The winding number for $E_{b}=0$ and the average fractal dimension $\overline{\Gamma}$ for different boundary condition as a function of $\lambda$ with $g=0.5$ and $L=610$. Eigenenergies in the complex plane at (c1) $\lambda=1$ and (d1) $\lambda=0.5$ for different boundary conditions. Distribution of the eigenstates with minimum real energy for (c2)(d2) OBCs and (c3)(d3) PBCs for $g=2$, where the fitting function is given by Eq.~\eqref{fit}. For plots (c)-(d), we set $L=144$ for OBCs and $L=2584$ for PBCs.}
\label{F1}
\end{figure}
Numerically, one can use the fractal dimension, which is defined as
\begin{equation}
\Gamma_{\beta}=-\lim_{L\rightarrow\infty}\frac{\ln\xi_{\beta}}{\ln L},
\end{equation}
where $\xi_{\beta}=\sum_{j=1}^{L}[|\psi_{j}(\beta)|^4/|\psi_{j}(\beta)|^2]$ is the inverse participation ratio and $\psi_{j}(\beta)$ is denoted as the amplitude for the $\beta$th eigenstate at the $j$th site. The $\Gamma=0$ ($\Gamma=1$) for the localized (extended) state, and $0<\Gamma<1$ for the critical state. Since we are discussing this in finite size, a convergence to $0$ ($1$) is sufficient to show that it is a localized (extended) state.

We show the average fractal dimension $\overline{\Gamma}$ ($=\sum_{\beta=1}^{L}\Gamma_{\beta}$) as a function of $\lambda$ and $g$ under PBCs in Fig.~\ref{F1}(a). The results indicate the localization transition point in $\lambda=e^{g}$, where all eigenstates transition from extended to localized states. When the system is in the extended phase ($\lambda=1$), as shown in Fig.~\ref{F1}(c1), it has a non-trivial topological point gap at PBCs, while the energy spectrum collapses to real spectrum at OBCs, which implies that the system has a non-Hermitian skin effect and the direction of the localization can be described by the winding number~\cite{ZGong2018,HJiang2019}. We apply a magnetic flux $\phi$ through a finite non-Hermitian ring and the hopping amplitudes are multiplied by $e^{\pm i\phi/L}$ under a specific choice of gauge, i.e., $t_{l}\rightarrow t_{l}e^{i\phi/L}$ and $t_{r}\rightarrow t_{r}e^{-i\phi/L}$. Therefore, the winding number can be defined as
\begin{equation}\label{omega}
\omega(E_{b})=\lim_{L\rightarrow\infty}\frac{1}{2\pi i}\int_{0}^{2\pi} d\phi\frac{\partial}{\partial \phi} \ln\left[\mathrm{ det}(H-E_{b})\right],
\end{equation}
where $E_{b}$ is the base point. When the winding number $\omega=1$ (-1), the eigenstates corresponding to $E_b$ with OBCs exhibit left (right) skin effect. We show the average fractal dimension $\overline{\Gamma}$ for different boundary condition and winding number $\omega$ for $E_b=0$ as a function of $\lambda$ with $g=0.5$ in Fig.~\ref{F1}(b), and one can see that before the Anderson localization ($\lambda<e^{g}$), the localization properties depend on boundary conditions, and the winding number signifies a skin phase. When $\lambda>e^{g}$, the system undergoes Anderson localization, $\overline{\Gamma}\rightarrow 0$ is independent of the boundary conditions, and $\omega=0$ indicates that the system is topological trivial. Note that, the Anderson transition and topological transition share the same critical point.

On the one hand, when $\lambda=1<e^{0.5}$, the eigenenergy for different boundary conditions is shown in Fig.~\ref{F1}(c1). Notably, it is highly dependent on boundary conditions and collapses into a curve with no area under OBCs, surrounded by the spectrum of the PBC. This is the main feature of the non-Hermitian skin effect~\cite{KZhang2020}. In addition, the winding number $\omega = 1$ indicates the left skin effect under OBCs, as shown in Fig.~\ref{F1}(c2), which can be described by $\psi\propto e^{-gj}$ and the eigenstates under the PBCs remain extended, as shown in Fig.~\ref{F1}(c3).

On the other hand, when $\lambda=2>e^{0.5}$, the system undergoes a localization transition. Fig.~\ref{F1}(d1) indicates that boundary conditions have negligible effects on eigenvalues, with a winding number $\omega=0$. The distribution of eigenstates with minimum real energy for different boundary conditions is shown in Figs.~\ref{F1}(d2)(d3), fitting well with Eq.~\eqref{fit}, and exhibiting different LEs on the left and right sides of the localized center.

However, solving LEs for localized states in non-reciprocal systems with either OBCs or PBCs relies on the fact that the Hamiltonian can be similarly transformed into reciprocal systems~\cite{HJiang2019,APAcharya2024,LZTang2021,XCai2021,XCai2022,ZHWang2021,SLJiang2023}. When the nearest-neighbor hopping contains disorder or quasiperiodic modulation, its similar transformation becomes extremely complicated. The recent studies reveal that it is possible to characterize the skin effect with disorder by modified generalized Brillouin zones~\cite{ZQZhang2023,HLiu2023}, while it is not yet known whether this is applicable for localized states.

\subsection{Asymmetric transfer matrix method}
Since the similarity transformation method is limited, we propose a new universal method called the ``Asymmetric transfer matrix method''. This method is used to determine the localization properties and non-Hermitian skin effects of non-reciprocal disordered systems by solving LEs. Specifically, the LE of a single-particle disordered system can be characterized by the corresponding transfer matrix, that is, the exponential divergence rate is calculated based on the recurrence relation of the amplitude of wave functions~\cite{Soukoulis1982,ESorets1991,PSDavids1995,XLuo2021,FKKunst2019}. For Hermitian cases, since the localized states have spatial inversion symmetry, one only need to perform transfer matrix analysis on one side of the wave function to obtain the corresponding LE (the LE of the other side is exactly the same) . However, for non-reciprocal systems, the exponential decay behavior of the two sides of the asymmetric localized states’ wave function is different, which means that the transfer matrix of both sides needs to be investigated to obtain their corresponding LEs. Therefore, when solving LEs for a non-reciprocal system, one need to use this new \textit{asymmetric transfer matrix method}. The eigenequation for non-reciprocal AA model reads
\begin{equation}
E\psi_{j}=t_{l}\psi_{j+1}+V_{j}\psi_{j}+t_{r}\psi_{j-1}.
\end{equation}
One can obtain the corresponding transfer matrix from left $(\psi_{j}, \psi_{j-1})^{T}$ to right $(\psi_{j+1}, \psi_{j})^{T}$ as
\begin{equation}\label{TLAAH}
\begin{pmatrix}
\psi_{j+1} \\
\psi_{j}
\end{pmatrix}=T_{l,j}
\begin{pmatrix}
\psi_{j} \\
\psi_{j-1}
\end{pmatrix},~T_{l,j}=\begin{pmatrix}
\frac{E-V_{j}}{t_{l}}  & -\frac{t_{r}}{t_{l}}\\
 1 &0
\end{pmatrix},
\end{equation}
as well as the transfer matrix from right $(\psi_{j+1}, \psi_{j})^{T}$ to left $(\psi_{j-1}, \psi_{j})^{T}$ as
\begin{equation}\label{TR}
\begin{pmatrix}
\psi_{j-1} \\
\psi_{j}
\end{pmatrix}=T_{r,j}
\begin{pmatrix}
\psi_{j} \\
\psi_{j+1}
\end{pmatrix},~T_{r,j}=
\begin{pmatrix}
\frac{E-V_{j}}{t_{r}}  & -\frac{t_{l}}{t_{r}}\\
 1 &0
\end{pmatrix}.
\end{equation}
The corresponding LEs satisfy
\begin{equation}
\begin{aligned}
&\gamma_{l}(E)=\lim_{L\rightarrow\infty}\frac{1}{L}\ln\left \| \prod_{j=1}^{L} T_{l,j} \right \|,\\
&\gamma_{r}(E)=\lim_{L\rightarrow\infty}\frac{1}{L}\ln\left \| \prod_{j=L}^{1} T_{r,j} \right \|,
\end{aligned}
\end{equation}
where $\prod_{j=1}^{L} T_{l,j}=T_{l,L}\dots T_{l,2}T_{l,1}$, and $\prod_{j=L}^{1} T_{r,j}=T_{r,1}\dots T_{r,L-1}T_{r,L}$. The symbol $\left \| \cdot  \right \| $ denotes the norm of $\prod_{j=1}^{L} T_{l,j}$ (or $\prod_{j=L}^{1} T_{r,j}$), which in fact takes the absolute value of the largest eigenvalue in the matrix. To calculate $\gamma_{l/r}(E)$, we apply Avila's global theory of one-frequency analytical $SL(2,\mathbb{C})$ cocycle~\cite{Avila2015}. The first step is to get the phase by means of complexification method, i.e., $\theta\rightarrow\theta+i\epsilon$. In the large $\epsilon$ limit, a direct calculation yields
\begin{equation}
\begin{aligned}
&T_{l,j}=\frac{e^{-i2\pi\alpha j+|\epsilon|}}{t_{l}}\begin{pmatrix}
-\lambda   & 0\\
 0 &0
\end{pmatrix}+o(1),\\
&T_{r,j}=\frac{e^{-i2\pi\alpha j+|\epsilon|}}{t_{r}}\begin{pmatrix}
-\lambda   & 0\\
0 &0
\end{pmatrix}+o(1).
\end{aligned}
\end{equation}
Then, one can get
\begin{equation}\label{epsilo}
\begin{aligned}
&\gamma_{l/r,\epsilon}(E)=\ln|\frac{\lambda}{t_{l/r}}|+|\epsilon|+o(1),\\
\end{aligned}
\end{equation}
Avila's global theory ensures $\gamma_{\epsilon}(E)$ is a piecewise continuous convex linear function with respect $\epsilon$~\cite{Avila2015}. In large $\epsilon$ limit, the slope of $\gamma_{\epsilon}$ is always 1, which means that when $\epsilon$ is large enough, $\gamma_{l/r,\epsilon} =\ln|\frac{\lambda}{t_{l/r}}|+|\epsilon|$. If the energy $E$ lies in the spectrum, one can get
\begin{equation}
\gamma_{l/r,\epsilon}(E)=\max\left\{\ln\left|\frac{\lambda}{t_{l/r}}\right|+|\epsilon|,0\right\},~\forall \epsilon\ge 0.
\end{equation}
Consequently, by setting $\epsilon=0$, one can obtain
\begin{equation}\label{LEAAHH}
\gamma_{l/r}(E)=\max\left\{\ln\left|\frac{\lambda}{t_{l/r}}\right|,0\right\}.
\end{equation}
The corresponding eigenstate of the eigenvalue $E$ is an Anderson localized state, if and only if the affine functions (in a neighborhood of $\epsilon=0$) $\gamma_l(E),~\gamma_{r}(E)>0$. Therefore, the localization transition depends on $\min\left\{\gamma_{l},\gamma_r\right\}$, which gives the transition point as $\lambda_{c}=\max\left\{t_{l},t_{r}\right\}$. For the non-reciprocal AA model, the LEs for the left and right sides of localization center are $\gamma_l=\ln|\lambda|-g$ and $\gamma_r=\ln|\lambda|+g$, which is consistent with the results from Eq.~\eqref{gammanrAA}. Form the previous similar transformation, it can be observed that before the localization transition, the strength of leftward (rightward) hopping is greater than that of rightward (leftward) when $g > 0$ ($g < 0$), resulting in a left (right) skin effect under OBCs. In the localized phase, the non-reciprocal hopping manifests as an asymmetric localized state, and the direction of the skin effect aligns with the side where the LE is smaller. When the two LEs are equal, the non-reciprocal strength corresponds to the inversion point of skin effect. We will discuss it in depth further in the following sections.

\section{Non-reciprocal off-diagonal AA model}\label{Sec.4}
In this section, we discuss the connection between the assymmetric LEs pair and localization transition (under PBCs) and skin effect (under OBCs) in non-reciprocal off-diagonal AA model.

\begin{figure*}[htbp]
\centering
\includegraphics[width=17cm]{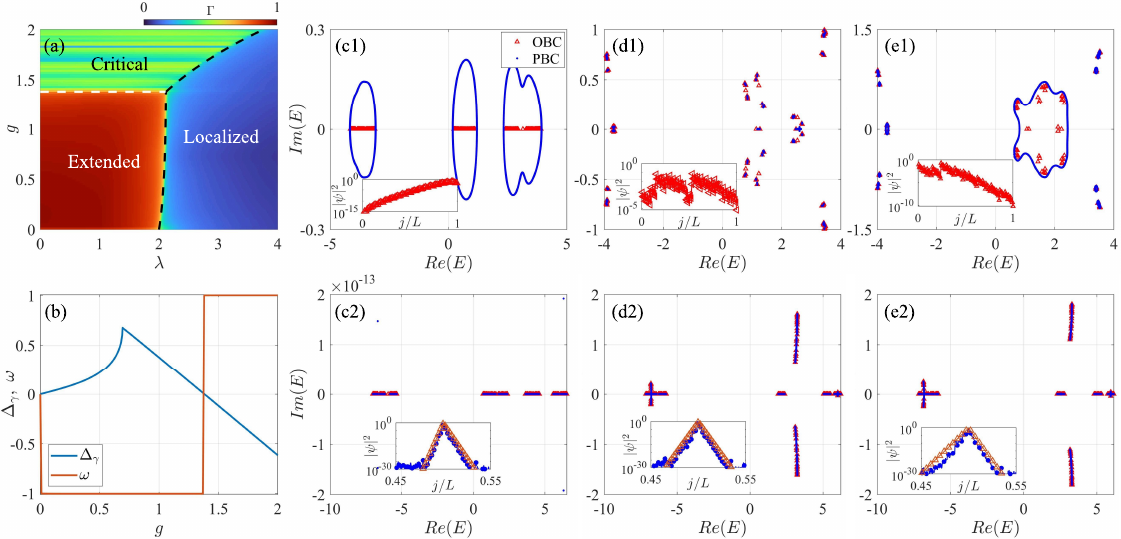}
\caption{(a) The average fractal dimension $\overline{\Gamma}$ in the $\lambda-g$ plane with $L=610$. (b) The differnet between the two sides of the LE $\Delta_{\gamma}$ and winding number $\omega$ as a function of $g$, where $E_{b}=0$ and $\lambda=0$ for $\omega$. Eigenenergies in the complex plane at (c1)-(e1) $\lambda=1$ and (c2)-(e2) $\lambda=3$ for OBCs (red triangle) and PBCs (blue dot), where the (c)-(e) plots correspond to $g=0.5$, $g=g_{c}$, and $g=1.6$, respectively. Insets in (c1)-(e1) [(c2)-(e2)] illustrate the distributions for OBCs (PBCs) of the eigenstate with the lowest real energy. For plots (c)-(e), we set $L=144$ for OBCs and $L=2584$ for PBCs. The orange lines are the analytical results from Eq.~\eqref{fit}.}
    \label{F3}
\end{figure*}
We consider the following Hamiltonian configured for Eq~\eqref{Hami}, i.e.,
\begin{equation}\label{OFAA}
\begin{aligned}
&t_{l,j}=t+t_{l}\cos[2\pi\alpha (j+\frac{1}{2})+\theta],\\
&t_{r,j}=t+t_{r}\cos[2\pi\alpha (j+\frac{1}{2})+\theta],\\
&V_{j}=2\lambda\cos(2\pi\alpha j+\theta),
\end{aligned}
\end{equation}
where $t_{l}=e^{g}$ and $t_{l}=e^{-g}$. When the parameter $g$ is set to zero, the Hamiltonian simplifies to the Hermitian off-diagonal AA model~\cite{FLiu2015}. It's interesting to note that unlike certain studies in the Ref.~\cite{LZTang2021,XCai2021}, where non-reciprocal modulation is applied to the entire hopping term, in this case, the modulation only affects the quasiperiodic part, which presents a challenge in terms of performing a similarity transformation due to the specific form of the modulation.

To obtain the corresponding LE, we write the eigenequation for the Eq.~\eqref{OFAA}
\begin{equation}
\begin{aligned}
E\psi_{j}=t_{l,j}\psi_{j+1}+t_{r,j-1}\psi_{j-1}+V_{j}\psi_{j},
\end{aligned}
\end{equation}
and the transfer matrices in both directions are
\begin{equation}
\begin{aligned}
&T_{l,j}=\begin{pmatrix}
\frac{E-V_{j}}{t_{l,j}}   & -\frac{t_{r,j-1}}{t_{l,j}} \\
1  & 0
\end{pmatrix}=A_{l,j}B_{l,j},\\
&T_{r,j}=\begin{pmatrix}
\frac{E-V_{j}}{t_{r,j-1}}   & -\frac{t_{l,j}}{t_{r,j-1}} \\
1  & 0
\end{pmatrix}=A_{r,j-1}B_{r,j},
\end{aligned}
\end{equation}
where
\begin{equation}
\begin{aligned}
&A_{l/r,j}=\frac{1}{t+t_{l/r}\cos\left[2\pi\alpha(j+\frac{1}{2})+\theta\right]},\\
&B_{l/r,j}=\begin{pmatrix}
E-V_{j}   & -t_{r/l,j-1/j} \\
t_{l/r,j/j-1}  & 0
\end{pmatrix}.
\end{aligned}
\end{equation}
The LE of the single-particle state is computed by
\begin{equation}
\gamma_{l/r}(E)=\lim_{L\rightarrow \infty}\frac{1}{L}\ln\left \|\prod_{j=1}^{L}T_{l/r,j}   \right \| =\gamma_{l/r}^{A}(E)+\gamma_{l/r}^{B}(E).
\end{equation}
Base on Weyl's equidistribution theorem and the properties of irrational rotation~\cite{HWeyl1916, GHChoe1993}, one can uniformly fill the interval $(0,2\pi]$ as $j$ varies. Then, by using the classical Jenson’s formula~\cite{SLonghi2019a,Gradshteyn2000}, we obtain $\gamma_{l/r}^{A}$ as
\begin{equation}
\begin{aligned}
\gamma_{l/r}^{A}(E)&=\lim_{L\rightarrow \infty}\frac{1}{L}\ln\left \|\prod_{j=1}^{L}\frac{1}{t+t_{l/r}\cos\left[2\pi\alpha(j+\frac{1}{2})+\theta\right]}   \right \|\\
&=\frac{1}{2\pi}\int_{0}^{2\pi} \ln\frac{1}{|t+t_{l/r}\cos(\theta)|}d\theta \\
&=\left\{
\begin{matrix}
\ln\frac{2}{t+\sqrt{t^2-t_{l/r}^2}},  &t\ge t_{l/r}, \\
\ln\frac{2}{t_{l/r}} & t< t_{l/r}.
\end{matrix}\right.
\end{aligned}
\end{equation}
For the $\gamma_{l/r}^{B}(E)$, we employ the Avila's global theory. First, we also need to get complexification of the phase, i.e., $\theta\rightarrow\theta+i\epsilon$. In the large $\epsilon$ limit, a direct calculation yields
\begin{equation}
B_{l/r,j}(\epsilon)=\frac{e^{-i2\pi\alpha j+|\epsilon|}}{2}
\begin{pmatrix}
-2\lambda  & -t_{l/r}e^{\mp i\pi\alpha}\\
-t_{r/l}e^{\pm i\pi\alpha}  & 0
\end{pmatrix}+o(1).
\end{equation}
Avila's global theory ensures $\gamma^{B}_{l/r}(\epsilon)$ is a continuous, convex, piecewise linear function with respect to $\epsilon$~\cite{Avila2015}. The corresponding LE is
\begin{equation}
\gamma_{l/r}^{B}=\ln\left|\frac{\lambda+\sqrt{\lambda^2-t_{l}t_{r}}}{2}\right|+\epsilon.
\end{equation}
Combined with $\gamma_{l/r}^{A}$ and let $\epsilon=0$, the LE is
\begin{equation}\label{Lc}
\gamma_{l/r}(E)=\left\{\begin{matrix}
 \max\left\{\ln\left|\dfrac{\lambda+\sqrt{\lambda^2-t_{l}t_{r}}}{t_{l/r}}\right|,0\right\}, &t< t_{l/r}, \\
 \max\left\{\ln\left|\dfrac{\lambda+\sqrt{\lambda^2-t_{l}t_{r}}}{(t+\sqrt{t^2-t_{l/r}^2})}\right|,0\right\},  &t\ge  t_{l/r}.
\end{matrix}\right.
\end{equation}
For the localized state should both $\gamma_{r/l}>0$, thus, the localization transition point is
\begin{equation}\label{lambdac}
\lambda_c=\left\{\begin{matrix}
\frac{t_{l}^2+1}{2t_{l}},  & t_l>t_r>t, \\
\max\left\{\frac{t_{l}^2+1}{2t_{l}},\frac{1+(t+\sqrt{t^2-t_{r}^2})^2}{2(t+\sqrt{t^2-t_{r}^2})}\right\},  & t_l>t>t_r, \\
\frac{1+(t+\sqrt{t^2-t_{r}^2})^2}{2(t+\sqrt{t^2-t_{r}^2})},  & t>t_l>t_r,\\
\end{matrix}\right.
\end{equation}
which depends on the one with the smallest LE. The first and third case correspond to smaller LEs on the left and right sides, respectively. However, for the second case an inversion of the asymmetric localized state occurs at the critical $g_{c}$, where the $g_{c}$ is given by $\gamma_l=\gamma_r$ and satisfies the following equation $e^{4g_{c}}-2te^{3g_{c}}+1=0$. Specifically, for $g<g_{c}$ ($g>g_{c}$), the $\gamma_{l}>\gamma_{r}$ ($\gamma_{l}<\gamma_{r}$) and the left (right) hand side of the localized state is more localized.

\subsection{Case $t=2$}
Let us start with $t=2$. To depict the phase diagram on the $\lambda-g$ plane, we introduce the average fractal dimension $\overline{\Gamma}=\frac{1}{L}\sum_{\beta=1}^{L}\Gamma_{\beta}$ and the results as shown in Fig.~\ref{F3}(a). Comparing with the previously analyzed Eq.~\eqref{lambdac}, we observe that when $g<g_c$ ($g>g_c$), $\lambda_c$ will act as a phase transition point from the extended phase (critical phase) to the localized phase, and the corresponding localized state has a smaller LE on the right (left) side. Interestingly, when both LEs are equal, the corresponding critical strength $g_{c}$ marks the phase transition point between the extended and critical phases.

Specifically, we assume that the system is in localized phase and define the difference between the two sides of the LE
\begin{equation}
\Delta_{\gamma}=\gamma_{l}-\gamma_{r}=
\left\{\begin{matrix}
\ln|\frac{t+\sqrt{t^2-t_{r}^2}}{t+\sqrt{t^2-t_{l}^2}}|,  & t>t_{l}>t_{r}, \\
\ln|\frac{t_{l}}{t+\sqrt{t^2-t_{l}^2}}|,  & t_{l}>t>t_{r},
\end{matrix}\right.
\end{equation}
which does not depend on the quasiperiodic strength $\lambda$. As shown by the blue line in Fig.~\ref{F3}(b), the LE on the left (right) side is larger when $g < g_{c}$ ($g>g_{c}$), while at the critical point $g = g_{c}$, the LE is the same on both sides. The transition of the LE is associated with the winding number under the delocalized phase. As shown by the orange line in Fig.~\ref{F3}(b), we have computed the corresponding winding number $\omega$ for $\lambda=0$ by using Eq.~\eqref{omega}, and the result shows that the system undergoes a topological phase transition at the transition point of the LE $g=g_{c}$, and the system under OBCs transitions from a right-skin phase to a left-skin phase. This means that the LE can predict the winding number of the delocalized phase and the direction of skin under OBCs on the smaller side.

To verify the above theoretical prediction, we discuss the cases $g = 0.5$, $g=g_{c}$ and $g = 1.5$, respectively [see Figs.~\ref{F3}(c)-(e)]. For $g<g_{c}$, when $\lambda=0.5$, the spectrum under PBCs forms multiple loops and includes the eigenstates under OBCs [see Figs.~\ref{F3}(c1)(c2)]. In the inset, we show the eigenstate distribution corresponding to the lowest real eigenenergy under OBCs with an exponential localization on the right boundary indicating a right-skin state. In Fig.~\ref{F3}(c2), we present the eigenenergy for different boundary conditions in the localized phase. Similar to the discussion of the non-reciprocal AA model, the eigenvalues are not affected by the boundary conditions. In the inset, we give the distributions of the lowest real eigenenergy for PBCs. It can be seen clearly that the numerical results are in good agreement with the analytical results (orange lines), the LE is smaller on the right side of the localization center.

When $g=g_{c}$, the localized state is symmetric and it is simultaneously the phase transition point of the extended-critical phase [see Figs.~\ref{F3}(d1)(d2)]. For $\lambda=0.5$, as shown in Fig.~\ref{F3}(d1), the eigenenergy does not form closed loops with area~\cite{KZhang2020}, which means no skin effect occurs in the system under OBCs. In the inset, by observing the distribution of eigenstates under OBC, one can see that the wave function is extended but not ergodic, which indicates that it is a critical state. The eigenenergy under the localized phase for $\lambda=3$ is shown in Fig.~\ref{F3}(d2), and the distribution of its eigenstates indicates that it is a symmetric localized state.

For the case $g = 1.5$ at $\lambda = 0.5$ [see Fig.~\ref{F3}(e)], the eigenvalues under PBCs similarly form multiple loops that encompass the eigenstates of OBCs and the eigenstate is left-skin state consistent with our theoretical analysis. When $\lambda = 3$, the system enters the localized phase, and its eigenenergy spectrum remains independent of the boundary conditions. Additionally, in the inset, we present the distribution of the eigenstates with the lowest eigenenergy. It is evident that the LE is smaller to the left of the localized center. This suggests that as the non-reciprocal quasiperiodic coupling $g$ strength increases, the skin direction reverses and the critical point is consistent with a inversion point of the LEs pair.

\begin{figure}[tbp]
\centering
\includegraphics[width=8.2cm]{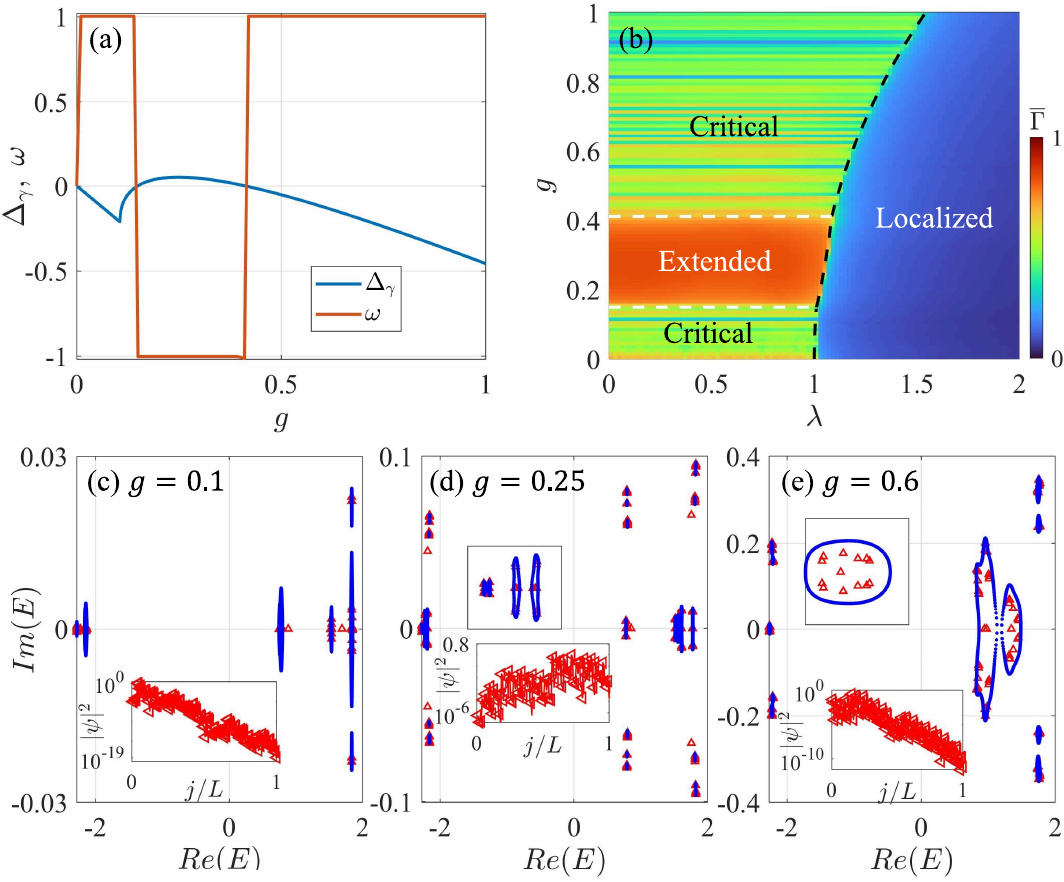}
\caption{(a) The different between the two sides of the LE $\Delta_{\gamma}$ and winding number $\omega$ as a function of $g$, where $E_{b}=0$ and $\lambda=0$ for $\omega$. (b) The average fractal dimension $\overline{\Gamma}$ in the $\lambda-g$ plane for $t=0.9$ with $L=610$. Eigenenergies in the complex plane at (c) $g=0.1$,(d) $g=0.25$, and (e) $g=0.6$ for different boundary conditions, where the inset in (c)-(e) shows the distribution of eigenstates with the minimum real eigenenergy. We set $\lambda=0.5$ for plots (c)-(e). The system size $L=144$ for OBCs and $L=2584$ for PBCs.}
    \label{F7}
\end{figure}

\subsection{Case $t=0.9$}
When $t=0.9$, it changes from $t_{l}>t_{r}>t$ to $t_{l}>t>t_{r}$ as $g$ increases to $g>-\ln(t)$. For case $t_{l}>t_{r}>t$, the left LE is smaller. Note that, for case $t_{l}>t_{r}>t$, there are two critical points $g_{c,1}=0.1493$ and $g_{c,2}=0.4114$. As shown in Fig.~\ref{F7}(a), we compare the difference between the two LEs and the winding number $\omega$. It can be seen that at $g<g_{c,1}$ and $g>g_{c,2}$, the LE on the left side is smaller, corresponding to $\omega=1$. When $g_{c,1}<g<g_{c,2}$, the LE on the right side is smaller, corresponding to $\omega=-1$. This indicates that there are two transitions in both the asymmetric localized state and the skin direction. The average fractal dimension $\overline{\Gamma}$ characterizes this reentrant phase transition well, as shown in Fig.~\ref{F7}(b). The localization transition can be described by Eq.~\eqref{lambdac}. However, for the delocalized phase, the $\overline{\Gamma}$ suggests the existence of a re-entrant critical phase, i.e., the system undergoes a critical-extended phase transition as $g$ increases, followed by a return to the critical phase once again. The corresponding critical points of phase transtion are consistesnt with topological phase transition points as well as the inversion points of asymmetric localized state.

Further, in Figs.~\ref{F7}(c)-(e), we plot the eigenenergy and the eigenstates' distribution for different $g$. The results show that the eigenstates transforms from left-skin to right-skin and finally back to left-skin under OBCs as $g$ grows. This indicates that asymmetric localized states can predict the skin direction of the skin phase.

\section{Non-reciprocal mosaic quasicrystal}\label{Sec.5}
Finally, we discuss a typical case with mobility edges, i.e., non-reciprocal mosaic AA model, where delocalized and localized states coexist. The corresponding Hamiltonian is as follows:
\begin{equation}\label{MEHami}
H=\sum_{j=1}^{L-1}[(t+W_{j})c_{j}^{\dagger}c_{j+1}+tc_{j+1}^{\dagger}c_{j}]+\sum_{j=1}^{L}V_{j}c_{j}^{\dagger}c_{j},
\end{equation}
where
\begin{equation}\label{MEHami}
\left\{W_{j},V_{j}\right\}=\left\{\begin{matrix}
\left \{W,2\lambda\right\}\cos(2\pi\alpha j+\theta),   &j=n\kappa, \\
0 , & \mathrm{else} .
\end{matrix}\right.
\end{equation}
Since the quasiperiodic potential periodically appears with an interval of $\kappa$, the system can be viewed as a quasicell of $\kappa$ sublattices, with the system having $N = L/\kappa$ quasicells and $n=1,2,\dots,N$ as the quasicell index, where $L$ is the system size. Unlike the standard mosaic model~\cite{YWang2020}, we introduce quasiperiodic hopping in the hopping from $n\kappa$-th lattice to $(n\kappa+1)-$th lattice with $W$ strength. We set $t=1$ as the unit energy.

One can get the transfer matrices of two direction as
\begin{equation}
\begin{aligned}
&T_{l,j}=\begin{pmatrix}
\frac{E-V_{j}}{1+W_{j}}  & -\frac{1}{1+W_{j}}  \\
 1 &0
\end{pmatrix}
\begin{pmatrix}
E  & -1  \\
 1 &0
\end{pmatrix}^{\kappa-1},\\
&T_{r,j}=\begin{pmatrix}
E  & -1  \\
 1 &0
\end{pmatrix}^{\kappa-1}
\begin{pmatrix}
E-V_{j}  & -1-W_{j}  \\
 1 &0
\end{pmatrix},
\end{aligned}
\end{equation}
where
\begin{equation}
\begin{pmatrix}
E  & -1  \\
 1 &0
\end{pmatrix}^{\kappa-1}=
\begin{pmatrix}
a_{\kappa}  & -a_{\kappa-1}  \\
a_{\kappa-1}  & -a_{\kappa-2}
\end{pmatrix},
\end{equation}
and the $a_{\kappa}$ is defined as
\begin{equation}
a_{\kappa}=\frac{1}{\sqrt{E^2-4}}\left[\left(\frac{E+\sqrt{E^2-4}}{2} \right)^{\kappa}-\left(\frac{E-\sqrt{E^2-4}}{2} \right)^{\kappa} \right].
\end{equation}

Similar to the previous calculation, by complexifying the phases $\theta\rightarrow\theta+i\epsilon$ and combining them with Avila's global theory we get~\cite{Avila2015}
\begin{equation}\label{MEL}
\gamma_{l}(E)=\left\{\begin{matrix}
\max\left\{\frac{1}{\kappa}\ln\left|\frac{2K_{\kappa}}{1+\sqrt{1-W^2}}\right|,0\right\}, &1\ge W, \\
\max\left\{\frac{1}{\kappa}\ln\left|\frac{2K_{\kappa}}{W}\right|,0\right\},  &1<  W,
\end{matrix}\right.
\end{equation}
and
\begin{equation}
\gamma_{r}(E)=\max\left\{\frac{1}{\kappa}\ln\left|K_{\kappa}\right|,0\right\},
\end{equation}
where $K_{\kappa}=\lambda a_{\kappa}+\frac{W}{2}a_{\kappa-1}$. Thus, the difference between the left and right LEs of the localized state is
\begin{equation}
\Delta_{\gamma}=\gamma_{l}-\gamma_{r}=
\left\{\begin{matrix}
\frac{1}{\kappa}\ln\left|\frac{2}{1+\sqrt{1-W^2}}\right|,  & 1\ge W, \\
\frac{1}{\kappa}\ln\left|\frac{2}{W}\right|,  & 1<W,
\end{matrix}\right.
\end{equation}
When $W=0$ and $W=2$, $\Delta_{\gamma}=0$ indicates that the localized state is symmetric, while at $0<W< 2$ ($W>2$), the right (left) LE is smaller.

For $\kappa=1$, $K_{1}=\lambda$ the system without mobility edge. The localization transition depend on the smaller LE, and the critical point is
\begin{equation}
\lambda_c=\left\{\begin{matrix}
1,  & W<2, \\
\frac{W}{2},  & W>2.\\
\end{matrix}\right.
\end{equation}
The result is similar to the previous discussion in that a phase transition of the extended-critical phase and an inversion of the asymmetric localized state occurs at $W = 2$. Also in the delocalized phase, there is a transition from a right skin to a left skin effect under OBCs.

For $\kappa=2$, $K_{2}=\lambda E+\frac{W}{2}$, the LE is energy dependent and the system has mobility edge. Since the system is non-Hermitian, delocalized states and localized states in the complex plane are separated by the mobility ring and correspond to the energy regions inside and outside the ring, respectively~\cite{LWang2024a,SZLi2024a,LWang2024b}. Comparing to $\gamma_{l}(E)$ and $\gamma_{r}(E)$, one can obtain the mobility rings for different $W$
\begin{equation}\label{MR}
\left\{\begin{matrix}
(E_{r}+\frac{W}{2\lambda})^2+E_{I}^2=(\frac{1}{\lambda})^2,  & W< 2,\\
(E_{r}+\frac{W}{2\lambda})^2+E_{I}^2=(\frac{W}{2\lambda})^2,  & W>2,
\end{matrix}\right.
\end{equation}
where $E_{r}$ and $E_{I}$ are the real and image eigenenergies, respectively. In terms of real energy, the corresponding mobility edge that separates the extended state from the critical state is located at $E_{c}$, i.e.,
\begin{equation}\label{ME}
E_{c}=\left\{\begin{matrix}
\frac{\pm 2-W}{2\lambda}, &W<2, \\
\frac{\pm W-W}{2\lambda},  & W>2.
\end{matrix}\right.
\end{equation}

\begin{figure}[tbhp]
\centering
\includegraphics[width=8.2cm]{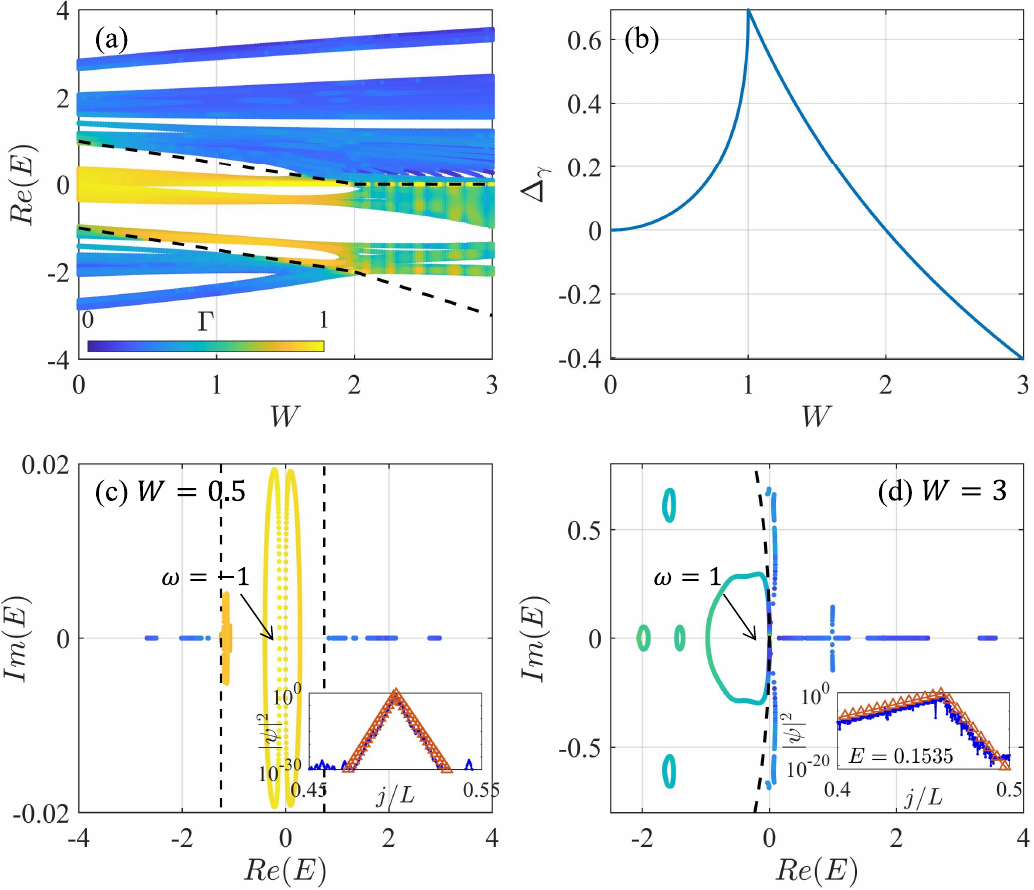}
\caption{(a) The fractal dimension of real eigenenergy corresponding eigenstates as a function of $W$ with the system size $L=610$, where the black dashed line is the mobility edge given by Eq.~\eqref{ME}. (b) The different between the two sides of the LE $\Delta_{\gamma}$ as a function of $g$. The fractal dimension $\Gamma$ of all eigenstates in complex plane for (c) $W=0.5$ and (d) $W=3$, where the black dashed line is the mobility ring given by Eq.~\eqref{MR}. Distribution of eigenstates with the inset (c) $E=2.678$, (d) $E=0.1535$ for PBCs. We set $t=1$, $\lambda=1$ for all plots and the system size $L=2584$ for plots (c)(d).}
    \label{F5}
\end{figure}

In Fig.~\ref{F5}(a), we show the fractal dimension $\Gamma$ as a function of $W$ for all eigenstates. Similar to what we have discussed earlier, $W=2$ is the phase transition point from the extended phase to the critical phase, which, however, also marks the inversion point of the asymmetric localized state. Comparing to the $\Delta_{\gamma}$ in Fig.~\ref{F5}(b), it can be seen that the transition between the two intermediate phases coincides with the transition point of the sign of $\Delta_{\gamma}$.

Further, we show the fractal dimension $\Gamma$ of all eigenstates in the complex plane for $W = 0.5$ and $W=3$ in Figs.~\ref{F5}(c) and (d), respectively. The black dashed line is the analytical mobility ring given by Eq.~\eqref{MR}. In the inset, we show the eigenstate distributions corresponding to the eigenstate with the smallest real eigenvalue and $E = 0.8048$ for $W = 0.5$ and $W =3$, respectively. It can be seen that the numerical results yield LEs for the localized states that are consistent with our theoretical analysis. Then we calculated the corresponding winding number for the reference energy $E_{b} = -0.2$, with $\omega= -1$ ($\omega=1$) at $W = 0.5$ ($W=3$) with respect to right (left) skin under OBCs, which agrees with LE predictions well. All the numerical and analytical evidence supports that the orientation of the smaller side of the localized state is consistent with the direction of skin effect as well as the spectral winding number.

\section{Conclusion}\label{Sec.6}

In summary, we have introduced an asymmetric transfer matrix method that effectively analyzes the characteristics of localization transitions and mobility edges in 1D non-reciprocal quasicrystals. This method proves to be efficient not only for conventional models that can be accurately studied using similar transformations but also for models where such transformations are not feasible.

To demonstrate the efficiency of this analytical approach, we initially apply it to the non-reciprocal AA model, a system that can also be solved using similar transformations. Upon comparing the outcomes derived from both methods, we observe a consistent agreement between the results obtained via the transfer matrix method and those obtained through the similar transformation method in this specific scenario.

Moreover, we apply this method to two classes of models that are not amenable to similar transformations, specifically, the off-diagonal non-reciprocal AA model (without a mobility edge) and the non-reciprocal Mosaic model (featuring a mobility edge). Differing from prior studies on systems with non-reciprocal disordered hopping~\cite{ZQZhang2023}, we investigate changes in the direction of skin modes not only through numerical techniques, but also by employing this newly introduced analytical approach. Specifically, for non-reciprocal off-diagonal systems, we not only demonstrate the inversion of the spatial distribution of the localized state wave function via precisely derived LEs but also accurately determine the direction of the skin modes. This method is equally applicable to cases involving mobility edges, such as non-reciprocal Mosaic AA models. The findings starkly reveal the inversion of the localized wave function and the explicit alteration in the direction of the skin modes. All numerical results presented are in agreement with the theoretical results derived from the asymmetric transfer matrix analysis.

\begin{acknowledgments}
This work was supported by the National Key Research and Development Program of China (Grant No.2022YFA1405300), the National Natural Science Foundation of China (Grant No.12074180), the Innovation Program for Quantum Science and Technology (Grant No. 2021ZD0301700), and the Guangdong Basic and Applied Basic Research Foundation (Grants No.2021A1515012350).
\end{acknowledgments}


\begin{thebibliography}{99}


\bibitem{PWAnderson1958} P. W. Anderson, Absence of diffusion incertain random lattices, Phys. Rev. \textbf{ 109}, 1492 (1958).


\bibitem{EAbrahams1979} E. Abrahams , P. W. Anderson, D. C. Licciardello and T. V. Ramakrishnan, Scaling Theory of Localization: Absence of Quantum Diffusion in Two Dimensions, Phys. Rev. Lett. \textbf{ 42}, 673 (1979).

\bibitem{PALee1985} P. A. Lee and T. V. Ramakrishnan, Disordered electronic systems, Rev. Mod. Phys. \textbf{ 57}, 287 (1985).

\bibitem{BHetenyi2021} B. Het\'enyi, S. Parlak, and M. Yahyavi, Scaling and renormalization in the modern theory of polarization: Application to disordered systems, Phys. Rev. B \textbf{104}, 214207 (2021).


\bibitem{FEvers2008} F. Evers and A. D. Mirlin, Anderson transitions, Rev. Mod. Phys. \textbf{ 80}, 1355 (2008).

\bibitem{ALagendijk2009} A. Lagendijk, B. Tiggelen, and D. S. Wiersma, Fifty years of Anderson localization, Phys. Today \textbf{ 62}, 24 (2009).






\bibitem{LFallani2007}  L. Fallani, J. E. Lye, V. Guarrera, C. Fort, and M. Inguscio, Ultracold atoms in a disordered crystal of light: Towards a Bose glass, Phys. Rev. Lett. \textbf{ 98}, 130404 (2007).

\bibitem{GRoati2008} G. Roati, C. D'Errico, L. Fallani, M. Fattori, C. Fort, M. Zaccanti, G. Modugno, M. Modugno, and M. Inguscio, Anderson localization of a non-interacting Bose-Einstein condensate, Nature (London) \textbf{ 453}, 895 (2008).

\bibitem{YLahini2009} Y. Lahini, R. Pugatch, F. Pozzi, M. Sorel, R. Morandotti, N. Davidson, and Y. Silberberg, Observation of a localization transition in quasiperiodic photonic lattices, Phys. Rev. Lett. \textbf{ 103}, 013901 (2009).

\bibitem{DTanese2014} D. Tanese, E. Gurevich, F. Baboux, T. Jacqmin, A. Lemaître, E. Galopin, I. Sagnes, A. Amo, J. Bloch, and E. Akkermans, Fractal energy spectrum of a polariton gas in a Fibonacci quasiperiodic potential, Phys. Rev. Lett. \textbf{ 112}, 146404 (2014).

\bibitem{FAAn2018} F. A. An, E. J. Meier, and B. Gadway, Engineering a fluxdependent mobility edge in disordered zigzag chains, Phys. Rev. X \textbf{ 8}, 031045 (2018).

\bibitem{HPLuschen2018} H. P. L\"uschen, S. Scherg, T. Kohlert, M. Schreiber, P. Bordia, X. Li, S. D. Sarma, and I. Bloch, Single-particle mobility edge in a one-dimensional quasiperiodic optical lattice, Phys. Rev. Lett. \textbf{ 120}, 160404 (2018).

\bibitem{VGoblot2020} V. Goblot, A. \ifmmode \check{S}\else \v{S}\fi{}trkalj, N. Pernet, J. L. Lado, C. Dorow, A. Lema\^{\i}tre, L. Le Gratiet, A. Harouri, I. Sagnes, S. Ravets, A. Amo, J. Bloch, and O. Zilberberg, Emergence of criticality through a cascade of delocalization transitions in quasiperiodic chains, Nat. Phys. \textbf{ 16}, 832 (2020).

\bibitem{FAAn2021}  F. A. An, K. Padavi\'e, E. J. Meier, S. Hegde, S. Ganeshan, J. H. Pixley, S. Vishveshwara, and B. Gadway, Interactions and mobility edges: Observing the generalized Aubry-Andr\'e model, Phys. Rev. Lett. \textbf{ 126}, 040603 (2021).

\bibitem{YWang2022} Y. Wang, J.-H Zhang, Y. Li, J. Wu, W. Liu, F. Mei, Y. Hu, L. Xiao, J. Ma, C. Chin, and S. Jia, Observation of interactioninduced mobility edge in an atomic Aubry-Andr\'e wire, Phys. Rev. Lett. \textbf{ 129}, 103401 (2022).

\bibitem{HLi2023} H. Li, Y.-Y Wang, Y.-H Shi, K. Huang, X. Song, G.-H Liang, Z.-Y Mei, B. Zhou, H. Zhang, J.-C Zhang, S. Chen, S.-P. Zhao, Y. Tian, Z.-Y Yang, Z. Xiang, K. Xu, D. Zheng, and H. Fan, Observation of critical phase transition in a generalized Aubry-Andr\'e-Harper model with superconducting circuits, npj Quantum Inf. \textbf{ 9}, 40 (2023).

\bibitem{SAubry1980} S. Aubry and G. Andr\'e, Analyticity breaking and Anderson localization in incommensurate lattices, Ann. Israel Phys. Soc \textbf{ 3}, 18 (1980).


\bibitem{JBiddle2010} J. Biddle and S. Das Sarma, Predicted Mobility Edges in One-Dimensional Incommensurate Optical Lattices: An Exactly Solvable Model of Anderson Localization, Phys. Rev. Lett. \textbf{ 104}, 070601 (2010).
\bibitem{JBiddle2011} J. Biddle, D. J. Priour, B. Wang, and S. Das Sarma, Localization in one-dimensional lattices with non-nearest-neighbor hopping: Generalized Anderson and Aubry-Andr\'e models, Phys. Rev. B \textbf{ 83}, 075105 (2011).
\bibitem{XDeng2019} X. Deng, S. Ray, S. Sinha, G. V. Shlyapnikov, and L. Santos, One-Dimensional Quasicrystals with Power-Law Hopping, Phys. Rev. Lett. \textbf{ 123}, 025301 (2019).
\bibitem{XXia2022} X. Xia, K. Huang, S. Wang, and X. Li, Exact mobility edges in the non-Hermitian ${t}_{1}\text{\ensuremath{-}}{t}_{2}$ model: Theory and possible experimental realizations, Phys. Rev. B \textbf{ 105}, 014207 (2022).

\bibitem{SRoy2021} S. Roy, T. Mishra, B. Tanatar, and S. Basu, Reentrant Localization Transition in a Quasiperiodic Chain, Phys. Rev. Lett. \textbf{ 126}, 106803 (2021).
\bibitem{MGoncalves2023} M. Gon\ifmmode \mbox{\c{c}}\else \c{c}\fi{}alves, B. Amorim, E. Castro, and P. Ribeiro, Phys. Rev. Lett. \textbf{ 131}, 186303 (2023).

\bibitem{SDasSarma1988} S. Das Sarma, S. He, and X. C. Xie, Mobility edge in a model one-dimensional potential, Phys. Rev. Lett. \textbf{ 61}, 2144 (1988).

\bibitem{SDasSarma1990} S. Das Sarma, S. He, and X. C. Xie, Localization, mobility edges, and metal-insulator transition in a class of one-dimensional slowly varying deterministic potentials, Phys. Rev. B \textbf{ 41}, 5544 (1990).

\bibitem{ZLu2022} Z. Lu, Z. Xu, and Y. Zhang, Exact Mobility Edges and Topological Anderson Insulating Phase in a Slowly Varying Quasiperiodic Model, Ann. Phys. (Berlin) \textbf{ 534}, 2200203 (2022).









\bibitem{APadhan2022} A. Padhan, M. Giri, S. Mondal, and T. Mishra, Emergence of multiple localization transitions in a one-dimensional quasiperiodic lattice, Phys. Rev. B \textbf{ 105}, L220201 (2022).

\bibitem{YWang2020} Y. Wang, X. Xu, L. Zhang, H. Yao, S. Chen, J. You, Q. Zhou, and X.-J. Liu, One-Dimensional Quasiperiodic Mosaic Lattice with Exact Mobility Edges, Phys. Rev. Lett. \textbf{ 125}, 196604 (2020).
\bibitem{SGaneshan2015} S. Ganeshan, J. H. Pixley, and S. D. Sarma, Nearest Neighbor Tight Binding Models with an Exact Mobility Edge in One Dimension, Phys. Rev. Lett. \textbf{ 114}, 146601 (2015).
\bibitem{HYao2019} H. Yao, H. Khouldi, L. Bresque, and L. Sanchez-Palencia, Critical behavior and fractality in shallow one-dimensional quasi-periodic potentials, Phys. Rev. Lett. \textbf{ 123}, 070405 (2019).
\bibitem{XLi2020} X. Li and S. Das Sarma, Mobility edge and intermediate phase in one-dimensional incommensurate lattice potentials, Phys. Rev. B \textbf{ 101}, 064203 (2020).

\bibitem{YCZhang2022} Y.-C. Zhang and Y.-Y. Zhang, Lyapunov exponent, mobility edges, and critical region in the generalized Aubry-Andr\'e model with an unbounded quasiperiodic potential, Phys. Rev. B \textbf{ 105}, 174206 (2022).
\bibitem{TLiu2022} T. Liu, X. Xia, S. Longhi, and L. Sanchez-Palencia, Anomalous mobility edges in one-dimensional quasiperiodic models, SciPost Phys. \textbf{ 12}, 027 (2022).
\bibitem{XCZhou2023} X.-C. Zhou, Y. Wang, T.-F. Poon, Q. Zhou, and X.-J. Liu, Exact new mobility edges between critical and localized states, Phys. Rev. Lett. \textbf{ 131}, 176401 (2023).
\bibitem{XPLi2016} X. P. Li, J. H. Pixley, D. L. Deng, S. Ganeshan, and S. Das Sarma, Quantum nonergodicity and fermion localization in a system with a single-particle mobility edge, Phys. Rev. B \textbf{ 93}, 184204 (2016).
\bibitem{XLi2017} X. Li, X. P. Li, and S. Das Sarma, Mobility edges in one-dimensional bichromatic incommensurate potentials, Phys. Rev. B \textbf{ 96}, 085119 (2017).




\bibitem{VVKonotop2016} V. V. Konotop, J. Yang, and D. A. Zezyulin, Nonlinear Waves in $\mathcal{PT}$-Symmetric Systems, Rev. Mod. Phys. \textbf{ 88}, 035002 (2016).

\bibitem{RElGanniny2018} R. El-Ganainy, K. G. Makris, M. Khajavikhan, Z. H. Musslimani, S. Rotter, and D. N. Christodoulides, Non-Hermitian Physics and PT Symmetry, Nat. Phys. \textbf{ 14}, 11 (2018).

\bibitem{YAshida2020} Y. Ashida, Z. Gong, and M. Ueda, Non-Hermitian physics, Adv. Phys. \textbf{ 69}, 249 (2020).

\bibitem{CMBender1998} C. M. Bender and S. Boettcher, Real Spectra in Non Hermitian Hamiltonians Having $\mathcal{PT}$ Symmetry, Phys. Rev. Lett. \textbf{ 80}, 5243 (1998).

\bibitem{AGuo2009} A. Guo, G. J. Salamo, D. Duchesne, R. Morandotti, M. Volatier-Ravat, V. Aimez, G. A. Siviloglou, and D. N. Christodoulides, Observation of $\mathcal{PT}$-Symmetry Breaking in Complex Optical Potentials, Phys. Rev. Lett. \textbf{ 103}, 093902 (2009).

\bibitem{ARegensburger2012} A. Regensburger, C. Bersch, M.-A. Miri, G. Onishchukov, D. N. Christodoulides, and U. Peschel, Parity-time synthetic photonic lattices, Nature (London) \textbf{ 488}, 167 (2012).

\bibitem{SWeimann2017} S. Weimann, M. Kremer, Y. Plotnik, Y. Lumer, S. Nolte, K. G. Makris, M. Segev, M. C. Rechtsman, and A. Szameit, Topologically protected bound states in photonic parity-time-symmetric crystals, Nat. Mater. \textbf{ 16}, 433 (2017).



\bibitem{MKremer2019} M. Kremer, T. Biesenthal, L. J. Maczewsky, M. Heinrich, R. Thomale, and A. Szameit, Demonstration of a two-dimensional $\mathcal{PT}$-symmetric crystal, Nat. Commun. \textbf{ 10}, 435 (2019).

\bibitem{SXia2021}  S. Xia, D. Kaltsas, D. Song, I. Komis, J. Xu, A. Szameit, H. Buljan, K. G. Makris, and Z. Chen, Nonlinear tuning of PT symmetry and non-Hermitian topological states, Science \textbf{ 372}, 72 (2021).

\bibitem{YLi2022} Y. Li, C. Liang, C. Wang, C. Lu, and Y.-C. Liu, Gain-Loss Induced Hybrid Skin-Topological Effect, Phys. Rev. Lett. \textbf{ 128}, 223903 (2022).

\bibitem{PPeng2016} P. Peng, W. Cao, C. Shen, W. Qu, J. Wen, L. Jiang, and Y. Xiao, Anti-Parity-Time Symmetry with Flying Atoms, Nat. Phys. \textbf{ 12}, 1139 (2016).

\bibitem{JLi2019} J. Li, A. K. Harter, J. Liu, L. de Melo, Y. N. Joglekar, and L. Luo, Observation of Parity-Time Symmetry Breaking Transitions in a Dissipative Floquet System of Ultracold Atoms, Nat. Commun. \textbf{ 10}, 855 (2019).

\bibitem{ZRen2022} Z. Ren, D. Liu, E. Zhao, C. He, K. K. Pak, J. Li, and G.-B. Jo, Chiral Control of Quantum States in Non-Hermitian Spin-Orbit-Coupled Fermions, Nat. Phys. \textbf{ 18}, 385 (2022).

\bibitem{LXiao2017} L. Xiao, X. Zhan, Z. H. Bian, K. K. Wang, X. Zhang, X. P. Wang, J. Li, K. Mochizuki, D. Kim, N. Kawakami, W. Yi, H. Obuse, B. C. Sanders, and P. Xue, Observation of Topological Edge States in Parity-Time-Symmetric Quantum Walks, Nat. Phys. \textbf{ 13}, 1117 (2017).

\bibitem{LLi2020} L. Li, C. H. Lee, S. Mu, and J. Gong, Critical non-Hermitian skin effect, Nat. Commun. \textbf{ 11}, 5491 (2020).
\bibitem{EJBergholtz2021} E. J. Bergholtz, J. C. Budich, and F. K. Kunst, Exceptional topology of non-Hermitian systems, Rev. Mod. Phys. \textbf{ 93}, 015005 (2021).

\bibitem{LZhou2023} L. Zhou and D. Zhang, Non-Hermitian Floquet topological matter—a review, Entropy \textbf{ 25}, 1401 (2023).

\bibitem{KKawabata2023} K. Kawabata, T. Numasawa, and S. Ryu, Entanglement phase transition induced by the non-hermitian skin effect, Phys. Rev. X \textbf{ 13}, 021007 (2023).

\bibitem{KLi2023} K. Li, Z.-C. Liu, and Y. Xu, Disorder-induced entanglement phase transitions in non-Hermitian systems with skin effects, arXiv:2305.12342.

\bibitem{LZhou2024} L. Zhou, Entanglement phase transitions in non-Hermitian quasicrystals, Phys. Rev. B \textbf{ 109}, 024204 (2024).
\bibitem{SZLi2024} S.-Z. Li, X.-J. Yu, and Z. Li, Emergent entanglement phase transitions in non-Hermitian Aubry-Andr\'e-Harper chains, Phys. Rev. B \textbf{ 109}, 024306 (2024).

\bibitem{HZLi2023} H.-Z. Li, X.-J. Yu, and J.-X. Zhong, Non-Hermitian stark manybody localization, Phys. Rev. A \textbf{ 108}, 043301 (2023).
\bibitem{XJYu2023} X.-J. Yu, Z. Pan, L. Xu, and Z.-X. Li, Non-Hermitian Strongly Interacting Dirac Fermions, Phys. Rev. Lett. \textbf{ 132}, 116503 (2024).
\bibitem{ZXGuo20222} Z.-X. Guo, X.-J. Yu, X.-D. Hu, and Z. Li, Emergent phase transitions in a cluster Ising model with dissipation, Phys. Rev. A \textbf{ 105}, 053311 (2022).

\bibitem{DWZhang2020} D.-W. Zhang, L.-Z. Tang, L.-J. Lang, H. Yan, and S.-L. Zhu, Non-hermitian topological anderson insulators, Sci. China Phys. Mech. \textbf{63}, 267062 (2020).

\bibitem{ZWang2022} Z. Wang, L.-J. Lang, and L. He, Emergent Mott insulators and non-Hermitian conservation laws in an interacting bosonic chain with noninteger filling and nonreciprocal hopping, Phys. Rev. B \textbf{105}, 054315 (2022).

\bibitem{CHXu2023} C.-H Xu, C.-Z Xu, Y.-Z Zhou, E.-H Cheng, and L.-J Lang, Electrical circuit simulation of non-Hermitian lattice models, Acta Phys. Sin. \textbf{72}, 200301 (2023).

\bibitem{JLi2023} J. Li, H.-T Ding, and D.-W Zhang, Quantum Fisher information and parameter estimation in non-Hermitian Hamiltonians, Acta Phys. Sin. \textbf{72}, 200601 (2023).


\bibitem{YXiong2018} Y. Xiong, Why Does Bulk Boundary Correspondence Fail in Some Non-Hermitian Topological Models?, J. Phys. Commun. \textbf{ 2}, 035043 (2018).

\bibitem{SYao2018} S. Yao and Z. Wang, Edge States and Topological Invariants of Non-Hermitian Systems, Phys. Rev. Lett. \textbf{ 121}, 086803 (2018).

\bibitem{ZGong2018}  Z. Gong, Y. Ashida, K. Kawabata, K. Takasan, S. Higashikawa, and M. Ueda, Topological Phases of Non-Hermitian Systems, Phys. Rev. X \textbf{ 8}, 031079 (2018).

\bibitem{VMMartinez2018} V. M. Martinez Alvarez, J. E. Barrios Vargas, and L. E. F. Foa Torres, Non-Hermitian robust edge states in one dimension: anomalous localization and eigenspace condensation at exceptional points, Phys. Rev. B \textbf{97}, 121401(R) (2018).



\bibitem{KZhang2020} K. Zhang, Z. Yang, and C. Fang, Correspondence Between Winding Numbers and Skin Modes in Non-Hermitian Systems, Phys. Rev. Lett. \textbf{ 125}, 126402 (2020).





\bibitem{KKawabata2019} K. Kawabata, K. Shiozaki, M. Ueda, and M. Sato, Symmetry and Topology in Non-Hermitian Physics, Phys. Rev. X \textbf{ 9}, 041015 (2019).
\bibitem{CHLee2019} C. H. Lee and R. Thomale, Anatomy of skin modes and topol ogy in non-Hermitian systems, Phys. Rev. B \textbf{ 99}, 201103(R) (2019).
\bibitem{KYokomizo2019} K. Yokomizo and S. Murakami, Non-Bloch Band Theory of Non-Hermitian Systems, Phys. Rev. Lett. \textbf{ 123}, 066404 (2019).
\bibitem{LXiao2020} L. Xiao, T. Deng, K. Wang, G. Zhu, Z. Wang, W. Yi, and P. Xue, Non-Hermitian bulk-boundary correspondence in quantum dynamics, Nat. Phys. \textbf{ 16}, 761 (2020).

\bibitem{NOkuma2020} N. Okuma, K. Kawabata, K. Shiozaki, and M. Sato, Topological Origin of Non-Hermitian Skin Effects, Phys. Rev. Lett. \textbf{ 124}, 086801 (2020).
\bibitem{DSBorgnia2020} D. S. Borgnia, A. J. Kruchkov, and R.-J. Slager, Non-Hermitian Boundary Modes and Topology, Phys. Rev. Lett. \textbf{ 124}, 056802 (2020).
\bibitem{ZYang2020} Z. Yang, K. Zhang, C. Fang, and J. Hu, Non-Hermitian Bulk Boundary Correspondence and Auxiliary Generalized Brillouin Zone Theory, Phys. Rev. Lett. \textbf{ 125}, 226402 (2020).
\bibitem{CXGuo2021} C.-X. Guo, C.-H. Liu, X.-M. Zhao, Y. Liu, and S. Chen, Exact Solution of Non-Hermitian Systems with Generalized Boundary Conditions: Size-Dependent Boundary Effect and Fragility of the Skin Effect, Phys. Rev. Lett. \textbf{ 127}, 116801 (2021).

\bibitem{LJLang2021} L.-J. Lang, Y. Weng, Y. Zhang, E. Cheng, and Q. Liang, Dynamical robustness of topological end states in non-reciprocal Su-Schrieffer-Heeger models with open boundary conditions, Phys. Rev. B \textbf{103}, 014302 (2021).



\bibitem{SLonghi2019} S. Longhi, Topological Phase Transition in Non-Hermitian Quasicrystals, Phys. Rev. Lett. \textbf{ 122}, 237601 (2019).


\bibitem{HJiang2019} H. Jiang, L.-J. Lang, C. Yang, S.-L. Zhu, and S. Chen, Interplay of non-Hermitian skin effects and Anderson localization in non-reciprocal quasiperiodic lattices, Phys. Rev. B \textbf{ 100}, 054301 (2019).



\bibitem{XCai2021} X. Cai, Boundary-dependent self-dualities, winding numbers, and asymmetrical localization in non-Hermitian aperiodic one-dimensional models, Phys. Rev. B \textbf{ 103}, 014201 (2021).

\bibitem{XCai2022} X. Cai, Localization transitions and winding numbers for non-Hermitian Aubry-Andr\'e-Harper models with off-diagonal modulations, Phys. Rev. B \textbf{ 106}, 214207 (2022).

\bibitem{LZTang2021} L.-Z. Tang, G.-Q. Zhang, L.-F. Zhang, and D.-W. Zhang, Localization and topological transitions in non-Hermitian quasiperiodic lattices, Phys. Rev. A \textbf{ 103}, 033325 (2021).
\bibitem{LZTang2022} L.-Z. Tang, S.-N. Liu, G.-Q. Zhang, and D.-W. Zhang, Topological Anderson insulators with different bulk states in quasiperiodic chains, Phys. Rev. A \textbf{ 105}, 063327 (2022).


\bibitem{YLiu2021} Y. Liu, Y. Wang, X.-J. Liu, Q. Zhou, and S. Chen, Exact mobility edges, $\mathcal{PT}$-symmetry breaking, and skin effect in one-dimensional non-Hermitian quasicrystals, Phys. Rev. B \textbf{ 103}, 014203 (2021).



\bibitem{YLiu2020} Y. Liu, X.-P. Jiang, J. Cao, and S. Chen, Non-Hermitian mobility edges in one-dimensional quasicrystals with parity-time symmetry, Phys. Rev. B \textbf{ 101}, 174205 (2020).

\bibitem{QBZeng2020a} Q.-B. Zeng, Y.-B. Yang, and Y. Xu, Topological phases in nonHermitian Aubry-Andr\'e-Harper models, Phys. Rev. B \textbf{ 101}, 020201(R) (2020).

\bibitem{QBZeng2020b} Q.-B. Zeng and Y. Xu, Winding numbers and generalized mobility edges in non-Hermitian systems, Phys. Rev. Res. \textbf{ 2}, 033052 (2020).

\bibitem{TLiu2020} T. Liu, H. Guo, Y. Pu, and S. Longhi, Generalized Aubry-Andr\'e self-duality and mobility edges in non-Hermitian quasi-periodic lattices, Phys. Rev. B \textbf{ 102}, 024205 (2020).

\bibitem{LJZhai2020} L.-J. Zhai, S. Yin, and G.-Y. Huang, Many-body localization in a non-Hermitian quasiperiodic system, Phys. Rev. B \textbf{ 102}, 064206 (2020).

\bibitem{YLiu2021a} Y. Liu, Q. Zhou, and S. Chen, Localization transition, spectrum structure and winding numbers for one-dimensional non-Hermitian quasicrystals, Phys. Rev. B \textbf{ 104}, 024201 (2021).



\bibitem{JClaes2021} J. Claes and T. L. Hughes, Skin effect and winding number in disordered non-Hermitian systems, Phys. Rev. B \textbf{ 103}, L140201 (2021).


\bibitem{SLonghi2021a} S. Longhi, Phase transitions in a non-Hermitian Aubry-Andr\'e-Harper model, Phys. Rev. B \textbf{ 103}, 054203 (2021).

\bibitem{SLonghi2021b} S. Longhi, Non-Hermitian Maryland model, Phys. Rev. B \textbf{ 103}, 224206 (2021).

\bibitem{LZhou2022} L. Zhou and Y. Gu, Topological delocalization transitions and mobility edges in the non-reciprocal Maryland model, J. Phys.: Condens. Matter \textbf{ 34}, 115402 (2022).

\bibitem{TQian2024} T. Qian, Y. Gu, and L. Zhou, Correlation-induced phase transitions and mobility edges in an interacting non-Hermitian quasicrystal, Phys. Rev. B \textbf{109}, 054204 (2024).



\bibitem{WHan2022} W. Han and L. Zhou, Dimerization-induced mobility edges and multiple reentrant localization transitions in non-Hermitian quasicrystals, Phys. Rev. B \textbf{ 105}, 054204 (2022).

\bibitem{QLin2022} Q. Lin, T. Li, L. Xiao, K. Wang, W. Yi, and P. Xue, Topological Phase Transitions and Mobility Edges in Non-Hermitian Quasicrystals, Phys. Rev. Lett. \textbf{ 129}, 113601 (2022).
\bibitem{DWZhang2020a} D.-W. Zhang, Y.-L. Chen, G.-Q. Zhang, L.-J. Lang, Z. Li, and S.-L. Zhu, Skin superfluid, topological Mott insulators, and asymmetric dynamics in an interacting non-Hermitian Aubry-Andr\'e-Harper model, Phys. Rev. B \textbf{101}, 235150 (2020).





\bibitem{NHatano1996} N. Hatano and D. R. Nelson, Localization transitions in non-hermitian quantum mechanics, Phys. Rev. Lett. \textbf{ 77}, 570 (1996).

\bibitem{APAcharya2024} A. P. Acharya, and S. Datta, Localization transitions in a non-Hermitian quasiperiodic lattice, Phys. Rev. B \textbf{ 109}, 024203 (2024).
\bibitem{ZHWang2021} Z.-H. Wang, F. Xu, L. Li, D. Xu, and B. Wang, Unconventional real-complex spectral transition and Majorana zero modes in non-reciprocal quasicrystals, Phys. Rev. B \textbf{ 104}, 174501 (2021).

\bibitem{SLJiang2023} S.-L. Jiang, Y. Liu, and L.-J. Lang, General mapping of one-dimensional non-Hermitian mosaic models to non-mosaic counterparts: Mobility edges and Lyapunov exponents, Chin. Phys. B \textbf{ 32}, 097204 (2023).



\bibitem{ZQZhang2023} Z.-Q. Zhang, H. Liu, H. Liu, H. Jiang, and X. C. Xie, Bulk-boundary correspondence in disordered non-Hermitian systems, Sci. Bull. \textbf{ 68}, 157 (2023).

\bibitem{HLiu2023} H. Liu, M. Lu, Z.-Q. Zhang, and H. Jiang, Modified generalized Brillouin zone theory with on-site disorder, Phys. Rev. B \textbf{ 107}, 144204 (2023).


\bibitem{Avila2015} A. Avila, Global theory of one-frequency Schr\"odinger operators, Acta Math. \textbf{ 215}, 1 (2015).

\bibitem{Soukoulis1982} C. M. Soukoulis and E. N. Economou, Localization in OneDimensional Lattices in the Presence of Incommensurate Potentials, Phys. Rev. Lett. \textbf{48}, 1043 (1982).


\bibitem{ESorets1991} E. Sorets and T. Spencer, Positive Lyapunov exponents for Schr\"odinger operators with quasi-periodic potentials, Commun. Math. Phys. \textbf{ 142}, 543 (1991).

\bibitem{PSDavids1995} P. S. Davids, Lyapunov exponent and transfer-matrix spectrum of the random binary alloy, Phys. Rev. B \textbf{ 52}, 4146 (1995).

\bibitem{XLuo2021} X. Luo, T. Ohtsuki, and R. Shindou, Transfer matrix study of the anderson transition in non-hermitian systems, Phys. Rev. B \textbf{104}, 104203 (2021).

\bibitem{FKKunst2019} F. K. Kunst and V. Dwivedi, Non-hermitian systems and topology: A transfer-matrix perspective, Phys. Rev. B \textbf{99}, 245116 (2019).

\bibitem{FLiu2015} F. Liu, S. Ghosh, and Y. D. Chong, Localization and adiabatic pumping in a generalized Aubry-Andr\'e-Harper model, Phys. Rev. B \textbf{ 91}, 014108 (2015).


\bibitem{HWeyl1916} H. Weyl, Ueber die Gleichverteilung von Zahlen mod. Eins, Math. Ann. \textbf{77}, 313 (1916).
\bibitem{GHChoe1993} G.H. Choe, Ergodicity and Irrational Rotations, Proc. Royal Irish Acad. A, \textbf{93A}, 193 (1993).


\bibitem{SLonghi2019a} S. Longhi, Metal-insulator phase transition in a non-Hermitian Aubry-Andr\'e-Harper model, Phys. Rev. B \textbf{ 100}, 125157 (2019).

\bibitem{Gradshteyn2000} I. S. Gradshteyn and I. M. Ryzhik, Table of Integrals, Series, and Products, 6th ed. (Academic Press, New York, 2000), integral number 4.226.

\bibitem{SZLi2024a} S.-Z. Li, and Z. Li, Ring Structure in the Complex Plane: A Fingerprint of non-Hermitian Mobility Edge, Phys. Rev. B \textbf{110}, L041102 (2024).

\bibitem{LWang2024a} L. Wang, Z. Wang, S. Chen, Non-Hermitian butterfly spectra in a family of quasiperiodic lattices, Phys. Rev. B \textbf{110}, L060201 (2024).

\bibitem{LWang2024b} L. Wang, J. Liu, Z. Wang, and S. Chen, Exact complex mobility edges and flagellate spectra for non-Hermitian quasicrystals with exponential hoppings, arXiv:2406.10769.






\end{thebibliography}
\end{document}